\documentclass[9pt, conference]{IEEEtran}
\IEEEoverridecommandlockouts

\usepackage{cite}
\usepackage{amsmath,amssymb,amsfonts}
\usepackage{algorithmic}
\usepackage{textcomp}
\usepackage{xcolor}
\usepackage{graphicx}
\usepackage{multirow}
\usepackage{makecell}
\usepackage{amssymb}
\usepackage{booktabs}
\usepackage{tablefootnote}
\usepackage{hyperref} 
\usepackage{bm}
\usepackage{caption} 
\usepackage{algorithm}
\usepackage{pifont} 
\usepackage{marvosym}

\makeatletter
\newcommand{\linebreakand}{%
  \end{@IEEEauthorhalign}
  \hfill\mbox{}\par
  \mbox{}\hfill\begin{@IEEEauthorhalign}
}
\makeatother

\def\BibTeX{{\rm B\kern-.05em{\sc i\kern-.025em b}\kern-.08em
    T\kern-.1667em\lower.7ex\hbox{E}\kern-.125emX}}
    
\begin{document}

\title{Breaking Through the Spike: Spike Window Decoding for Accelerated and Precise Automatic Speech Recognition}
\author{\IEEEauthorblockN{Wei Zhang$^{1, *}$, Tian-Hao Zhang$^{2, *}$\thanks{* Equal contribution.}, Chao Luo$^{1}$, Hui Zhou$^{1}$, Chao Yang$^{1}$,
Xinyuan Qian$^{2,}\textsuperscript{\Letter}$\thanks{\textsuperscript{\Letter} Corresponding author.}, Xu-Cheng Yin$^{2}$}
\IEEEauthorblockA{$^{1}$TRIP.COM GROUP, Shanghai, China\\
$^{2}$School of Computer and Communication Engineering, \\University of Science and Technology Beijing, Beijing, China\\
wzhangs@trip.com, tianhaozhang@xs.ustb.edu.cn}
}

\maketitle
\begin{abstract}
Recently, end-to-end automatic speech recognition has become the mainstream approach in both industry and academia. 
To optimize system performance in specific scenarios, the Weighted Finite-State Transducer (WFST) is extensively used to integrate acoustic and language models, leveraging its capacity to implicitly fuse language models within static graphs, thereby ensuring robust recognition while also facilitating rapid error correction.
However, WFST necessitates a frame-by-frame search of CTC posterior probabilities through autoregression, which significantly hampers inference speed.
In this work, we thoroughly investigate the spike property of CTC outputs and further propose the conjecture that adjacent frames to non-blank spikes carry semantic information beneficial to the model. Building on this, we propose the Spike Window Decoding algorithm, which greatly improves the inference speed by making the number of frames decoded in WFST linearly related to the number of spiking frames in the CTC output, while guaranteeing the recognition performance.
Our method achieves SOTA recognition accuracy with significantly accelerates decoding speed, proven across both AISHELL-1 and large-scale In-House datasets, establishing a pioneering approach for integrating CTC output with WFST.
\end{abstract}


\section{Introduction}
Automatic speech recognition (ASR) technology has seen remarkable growth in recent years\cite{Sainath2015, Dong2018, Wu2023,zhang24q_interspeech, zhou2024knn}. 
Recently, contemporary end-to-end (E2E) ASR systems \cite{Prabhavalkar2023} have gained popularity, surpassing traditional methods \cite{Juang1991, Povey2016} by eliminating the need for iterative alignment and requiring only a single model to achieve superior accuracy, speed, and training simplicity. E2E systems are typically categorized into CTC-based \cite{Graves2006, Amodei2016}, RNN-T-based \cite{Graves2012, Graves2014}, and AED-based models \cite{Chan2016, Zhang2023, DBLP:conf/interspeech/ZhangQLC00QY23}, as well as hybrid approaches that couple these methods, such as CTC/RNN-T and CTC/AED ASR systems \cite{Watanabe2017, Kim2017,  DBLP:conf/icassp/ZhangZZZL24}.

While E2E ASR systems excel in general-purpose applications, relying solely on an acoustic model often proves inadequate in specialized or vertical scenarios. In these scenarios, an additional language model is typically used to refine the final recognition output. A common approach involves generating N-best results from the acoustic model and applying shallow fusion with the language model to determine the highest-scoring sequence \cite{Chorowski2017, Kannan2018}. Previous studies also investigated generating N-best results from the CTC encoder, followed by non-autoregressive combination in the decoder to achieve optimal outcomes \cite{Sainath2021}. 
WFST \cite{Mohri2000, hori2022speech, DBLP:conf/icassp/0001CXP0K21} is extensively utilized in industrial ASR systems due to its high interpretability and scalability. EESEN \cite{Miao2015} pioneers the integration of E2E ASR with WFST, introducing the CTC-TLG paradigm, which using the CTC model to predict phone posteriors at each frame, followed by Viterbi beam search on a modified WFST network, achieving comparable performance with traditional HMM systems.

However, it is impractical to directly decode the complete CTC posterior output with the WFST due to the significant increasing of decoding latency. Extensive prior researches have been conducted to address this issue. Chen et al \cite{Chen2016} proposed the CTC Lattices based Phone Synchronous Decoding (LSD) algorithm which removed the redundancy caused by blank frames in the CTC output. Speech-LLaMA \cite{Dong2023} indicated that the blank frames in the CTC output still encode pertinent information, and an averaging technique was applied to these blank frames.  PolyVoice \cite{Wu2023} employed a method that systematically discards all blank frames. In this paper, we posit that blank frames in close proximity to non-blank spikes harbor the most essential semantic information. We assert that selectively incorporating a limited subset of these neighboring blank frames into the WFST decoding represents an optimal strategy for enhancing both performance and decoding efficiency. Building on this hypothesis, we introduce the Spike Window Decoding (SWD) algorithm, which leverages our novel spike window function to construct sequence of windowed neighborhoods centered around the non-blank spikes. Compared to the LSD and Discarding algorithm, our method not only acknowledges but also empirically validates the beneficial quality of the information encoded in blank frames adjacent to CTC spikes, thereby substantially elevating the system's performance ceiling. Additionally, unlike the vanilla density strategy and the averaging strategy, which utilize all blank frames, the SWD algorithm adeptly mitigates the adverse effects of redundant blank frames that contain non-contributory information. In summary, we present a novel and state-of-the-art (SOTA) paradigm for utilizing CTC outputs in the WFST decoding, which strategically harnesses the intrinsic properties of CTC spikes to attain exceptional performance while preserving decoding efficiency.


We present a comprehensive evaluation of our proposed SWD algorithm through extensive experiments conducted on the widely utilized AISHELL-1 \cite{Bu2017} Mandarin dataset, as well as a large-scale In-House dataset comprising 43000 hours of speech data. To establish a rigorous ASR baseline, we first develop a CTC/AED-based acoustic model and a GPU-optimized WFST \cite{Galvez2023}, achieving results that set a new benchmark for SOTA performance. Subsequently, we thoroughly explore the SWD algorithm, investigating its application to speech frames with left-neighboring, right-neighboring, and window-neighboring configurations. As a result, our method achieves Character Error Rates (CER) of 3.89\% and 2.09\% on the AISHELL-1 and In-House datasets, respectively, surpassing both baseline models and previous SOTA approaches in recognition accuracy. Concurrently, the proposed SWD algorithm significantly enhances inference speed, with improvements of 1.76 and 2.17 times over the baseline method for the respective datasets. This illustrates that our method achieves a remarkable balance between superior recognition performance and enhanced decoding speed across datasets of different scales. It robustly establishes our approach as a pioneering paradigm for decoding WFST with CTC outputs, advancing the SOTA in this domain.

\section{Related Work}
\label{sec:related work}
\subsection{Hybrid CTC/AED Model}
Given an input speech feature $\mathbf{X}$, we use $\mathbf{Y}$ to represent the label sequence corresponding to $\mathbf{X}$, which a length of $\mathrm{L}$. The jointly loss of hybrid CTC/AED E2E ASR system are provided as:
\begin{align}
    \label{eq1}
    &\mathbf{H}_\mathrm{encoder} = \mathrm{Encoder}(\mathbf{X}) \\
    &\mathbf{H}_\mathrm{decoder} = \mathrm{Decoder}(\mathbf{H}_\mathrm{encoder}, \mathbf{Y}) \\
    &\mathcal{L}_\mathrm{CTC}=\mathrm{CTC}(\mathbf{H}_\mathrm{encoder}, \mathbf{Y})\\
    &\mathcal{L}_\mathrm{AED}=\mathrm{CrossEntropy}(\mathbf{H}_\mathrm{decoder}, \mathbf{Y})\\
&\mathcal{L}=\alpha*\mathcal{L}_\mathrm{CTC} + (1-\alpha)*\mathcal{L}_\mathrm{AED}
\end{align}
where we use $\mathrm{T}$ to denote the length of $\mathbf{H}_\mathrm{encoder}$, and $\alpha$ is a hyper-parameter used to adjust the weight ratio between the encoder and decoder, with its value ranging from $[0,1]$, which is routinely configured to $0.1$ for the rest of the study.

\subsection{WFST based decoding algorithm}
In this work, the WFST system is integrated on the encoder side, culminating in the construction of the TLG graph. The process of constructing the static graphs is as follows:
\begin{align}
    \label{eq2}
    \mathbf{TLG} = \mathrm{T}\circ\mathrm{min}(\mathrm{det}(\mathrm{L}\circ\mathrm{G}))
\end{align}
where $\mathrm{L}$ and $\mathrm{G}$ denotes  $\mathrm{Lexicon.fst}$ and $\mathrm{Grammar.fst}$, respectively; $\mathrm{min}$,  $\mathrm{det}$ and $\circ$ are short for minimization, determinization and composition WFST operations. It must be spcified that we adopt GPU-based WFST decoding algorithm and the compact Token.fst (denoted as $\mathrm{T}$) from \cite{Galvez2023} which ultimately reduces the construction of state complexity from  second-order exponential level to linear level while maintaining  comparable accuracy.

\section{Methodology}
\label{sec:method}
Our proposed SWD algorithm first retrieves the indexes of the frames with the highest probabilities in the CTC posteriori probability matrix through the index function $\mathrm{argmax}$, and then obtain the indexes corresponding to the non-blank spike frames through the comparison function. Subsequently, the window sequence of neighboring frames is obtained via the innovative Spike Window function. The final decoded sequence is derived from the processed window sequence. Additionally, we employ a weight pushing strategy between $\mathrm{det}$ and $\mathrm{min}$ during the TLG WFST graph construction. The following sections will provide a detailed explanation of the SWD process.

\begin{figure}[t!]
  \centering
  \vspace{1mm}
  \includegraphics[width=0.85\linewidth]{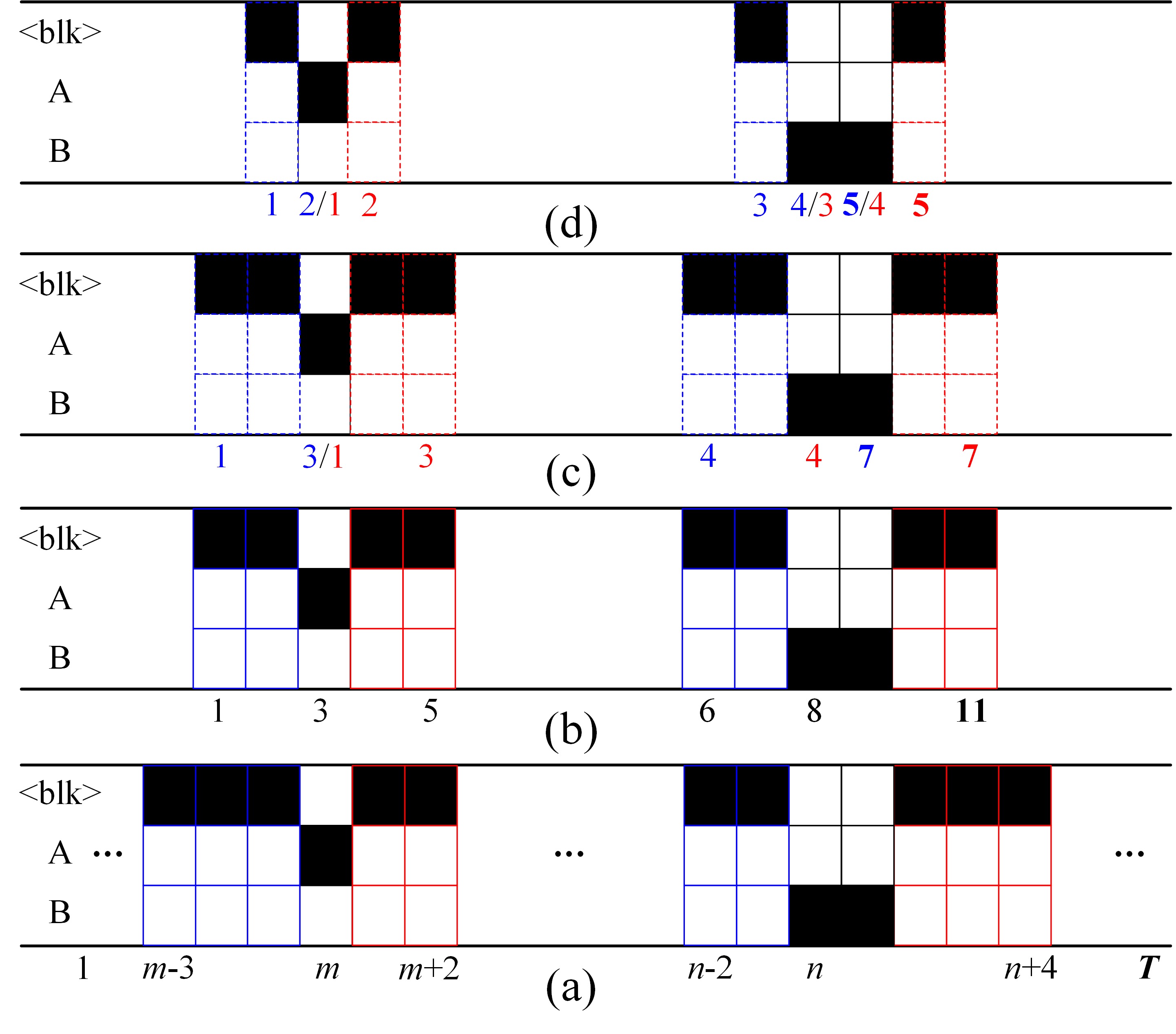}
  \caption{(a) represents a search performed using the dense frames, (b) and (c) indicate a neighboring frames of 2, while (d) indicates that the neighboring frame is 1.The dashed blue line signifies that only the left side is neighboring, whereas the red line denotes that only the right side is neighboring. The solid blue and red lines indicate that the left and right non-blank sides are neighboring simutaneously.}
  \label{fig:SWD}
\end{figure}

\subsection{Spike window decoding}
\label{sec:refs}
Drawing on the insight that the frames adjacent to CTC spikes encapsulate crucial information, we introduce the SWD algorithm. Initially, $\mathbf{H}_\mathrm{logits}$ is derived by applying $\mathrm{log\_softmax}$ to $\mathbf{H}_\mathrm{encoder}$ of Eq. \ref{eq1}. The index function $\mathrm{argmax}$ is then employed to identify the index of the maximum value within the posterior probability matrix, followed by calculating the non-zero maximum sequence. Here, $\mathbf{0}$ denotes the index corresponding to the modeling unit \textit{blank}, while the remaining indices correspond to the non-blank modeling units represented by $\mathbf{Y}_{\mathrm{logits\_spike}}$. The calculation process for $\mathbf{Y}_{\mathrm{logits\_spike}}$ is detailed as follows:
\begin{align}
    \label{eq3}
    \mathbf{Y}_{\mathrm{logits\_spike}}=\mathrm{argmax}(\mathbf{H}_{\mathrm{logits}}) \ != 0
\end{align}
thereafter, to obtain the sequence of window neighboring frames at spike positions, we introduce the spike window function $\mathrm{F_{sw}}$. The calculation principles underlying this function are presented as:
\begin{align}
    \label{eq4}
   &\mathbf{H}_{\mathrm{logits\_sw}} = \mathrm{F_{sw}}(\mathbf{Y}_{\mathrm{logits\_spike}}) \\
    &\mathrm{F_{sw}} = \mathrm{Concat}(\mathbf{H}_\mathrm{s}, \mathbf{H}_\mathrm{s \pm 1},...,\mathbf{H}_\mathrm{s \pm w}), \quad \mathbf{s} \in \mathbf{Y}_{\mathrm{logits\_spike}}
\end{align}
where $\mathbf{H}_\mathrm{s \pm w}$ represents the $\mathrm{w}$-th frame to the left and right of the spliced s-frame, and $\mathrm{w}$ denotes the range of the spliced frames. In order to get the final input $\mathbf{H}_\mathrm{logits\_final}$ to WFST , we propose the $\mathrm{F_{post}}$ function, which operates as follows: Initially, remove the frames in $\mathbf{H}_{\mathrm{logits\_sw}}$ which exceed the upper and lower limits of the numbering range of $\mathbf{Y}_{\mathrm{logits\_spike}}$; Subsequently, duplicate elements are eliminated and the remaining elements are sorted; Finally, mapping the sequence of frames after window neighboring to the corresponding $\mathbf{H}_{\mathrm{logits}}$ and obtaining $\mathbf{H}_{\mathrm{logits\_final}}$. Fig. \ref{fig:SWD} illustrates the schematic of the SWD algorithm, with black blocks representing the window sequence in the final SWD output. The algorithm effectively leverages the blank frames surrounding non-blank spikes, optimizing their contribution to the decoding process.

Conventional algorithms for WFST decoding precedure on CTC outputs require iterating over all frames in an autoregressive mode, wherein the states involved in WFST computation span the entire frame sequence, which leads to slow decoding speeds. However, SWD algorithm reduces the number of frames involved in WFST decoding procedure to be linearly proportional to the number of non-blank spikes in the CTC outputs. The theoretical proof proceeds as follows: First, we assume that all non-blank spikes are non-overlapping and that any two non-blank spikes are separated by more than $\mathrm{K} \times \mathrm{W}$ frames. Second, we assume that the distance of the first and last non-blank spike frames from the $0$-th and $\mathrm{T}$-th frames of the CTC outputs is less than the $\mathrm{W}$ frames. So that we can get:
\begin{equation}
    \mathrm{L}_{\mathrm{SWD}} = \mathrm{N} * (\mathrm{K} * \mathrm{W} + 1)
\end{equation}
where $\mathrm{L}_{\mathrm{SWD}}$ represents the total number of frames after applying the SWD algorithm, $\mathrm{N}$ denotes the total number of non-blank spikes, and $\mathrm{K}$ denotes the window coefficient, where $\mathrm{K}=1$ accounts for either the left or right frame, and $\mathrm{K}=2$ accounts for both left and right frames. Given that $\mathrm{K} \times \mathrm{W}$ is set to a small value and $\mathrm{N}$ is much smaller than $\mathrm{T}$, the output length of SWD is significantly reduced compared to the original CTC output. It is true that in practice there are cases where SWD does not fulfill the above two assumptions, such as in Fig. \ref{fig:SWD} where the two non-blank spikes are adjacent to each other, but in this case $\mathrm{L}_{\mathrm{SWD}}$ will only be smaller, since this means that fewer blank frames are considered. This explains why the inference efficiency is enhanced by the proposed SWD algorithm.
\vspace{-2mm}

\subsection{TLG graph optimization}
\label{TLG}
The major method of constructing CTC TLG graphs in existing work is depicted as Eq. \ref{eq2}. However, when dealing with large-scale language models, the constructed TLG graph will become significantly large, leading to increased memory usage, storage space, and a more complex inference process. To mitigate these issues, this work introduces the weight pushing operation to shift the weights of the corresponding sequence forward, which is applied between the $\mathrm{det}$ and $\mathrm{min}$ operation. This method allows for the pruning of low-probability paths in advance without compromising accuracy, thereby improving both the search space and speed. The construction process of static graphs is as follows:
\begin{align}
    \label{eq5}
   \mathbf{TLG} = \mathrm{T}\circ\mathrm{min}(\mathrm{push}({\mathrm{det}}(\mathrm{L}\circ\mathrm{G})))
\end{align}

\section{Experiment}
\label{experimet}
\vspace{-2mm}

\subsection{Dataset}
\label{dataset}
We conduct experiments on AISHELL-1 and a large-scale In-House dataset, on which the AISHELL-1 dataset is a 178-hour Mandarin dataset with a sampling rate of 16000 Hz. Our In-House dataset is also a Mandarin dataset which sampling rate is 8000 Hz, on which the training set contains about 43000 hours of labeled speech data, and the test set includes about 20 hours of labeled data for hundreds of vertical scenes.
\vspace{-1mm}

\subsection{Model architecture}
\vspace{-1mm}
\subsubsection{Acoustic model settings}
The acoustic model constructed in this work is a multi-task Hybrid CTC/AED structure, in which the encoder and decoder models are based on the Zipformer and transformer models, respectively. On the AISHELL-1 dataset, the encoder follows the medium size Zipformer in \cite{Yao2023}, the decoder side is based on the transformer standard model, which has 6 transformer blocks. For experiments on the In-Hourse dataset, the decoder is as the same as the one set in the AISHELL-1 dataset; for the encoder side, we use a configuration with a larger number of parameters. The final sizes of the acoustic models constructed on the AISHELL-1 and In-House datasets are 155.9M and 316.7M, respectively, and we will refer to the larger model as -XL in the following expriments.

\subsubsection{Language model settings}
For the AISHELL-1 dataset, the language model is constructed using only its 120098 training sets, and its 3-gram, 5-gram, and 7-gram language models are trained using the Srilm tool \footnote{https://www.sri.com/platform/srilm}. On the In-House dataset, we use about 1.5 billion pieces of Mandarin dialog text data to construct the language model. The operations $\mathrm{composition}$, $\mathrm{det}$, $\mathrm{min}$ and $\mathrm{weight \ pushing}$ introduced in Section \ref{sec:method} are implemented using the Openfst tool \footnote{https://www.openfst.org/twiki/bin/view/FST/WebHome}.

\begin{table}[!t]
\vspace{1mm}
    \caption{Results of baseline models on the AISHELL-1 dataset. $\dagger$ indicates the results presented in the paper, and $\ddagger$ stands for the results achieved in the ESPnet \cite{espnet}.}
    \centering
    \label{table1}
    \setlength{\tabcolsep}{2mm}{
    \begin{tabular}{l c cc}
    \toprule
    \multirow{2}{*}{\textbf{Model}} & \multirow{2}{*}{\textbf{Size (M)}}  & \multicolumn{2}{c}{\textbf{CER (\%)}} \\
    & & \makecell{Dev} & \makecell{Test}\\
    \midrule
    Conformer $\ddagger$ (CTC/AED) \cite{Gulati2020} & 46.2 & 4.5 & 4.9   \\
    E-Branchformer $\ddagger$ (CTC/AED) \cite{Kim2023} & 37.9 & 4.2 &  4.5   \\
    Branchformer $\ddagger$ (CTC/AED) \cite{Peng2022} & 45.9 & 4.19 & 4.43 \\
    Paraformer $\dagger$ (AED) \cite{DBLP:conf/interspeech/GaoZ0Y22} & 46 & 4.7 & 5.1 \\
    CIF-Transducer $\dagger$ (CTC/RNN-T) \cite{DBLP:conf/icassp/ZhangZZZL24}
& 130 & 4.1 & 4.3 \\
    Zipformer $\dagger$ \cite{Yao2023} (RNN-T) & 157.3 & 4.03 & 4.28 \\
    \midrule
    \textbf{A1} (CTC/AED Zipformer) & 155.9 & \textbf{3.98} & \textbf{4.19} \\
    \bottomrule
    \end{tabular}}
    \vspace{-4mm}
\end{table}

\subsubsection{Training, inference and evaluation}
During the training stage, 80-dimensional filter banks are extracted as speech features, with a frame length of 25 ms and a frame shift of 10 ms. To augment the data, a speech speed perturbation \cite{sp} is used, using perturbation coefficients of 0.9, 1.0, and 1.1. Furthermore, the SpecAugment \cite{spec} strategy is also used to enhance the robustness of the model. All models are trained on 16 NVIDIA Tesla A800 GPUs (80G) with mixed precision training. For inference, all our computations are consistently performed on a Tesla T4 GPU (16GB) with a 16-core CPU and 32GB of RAM. In order to achieve competitive results, AISHELL-1 and In-House dataset are trained for 200 epochs and 20 epochs, respectively. For the inference stage, we evaluate the recognition performance using the standard CER which is calculated by determining the Levenshtein distance \cite{levenshtein1966binary} between the labeled sequence and the predicted sequence.

\begin{table*}[!th]
    \caption{Experiment results on the AISHELL-1 dataset. The arrows indicate whether higher or lower values are preferable.}
    \centering
    \renewcommand{\arraystretch}{1.0}
    \label{tabel2}
    \setlength{\tabcolsep}{6.5mm}{
    \begin{tabular}{cc c ccc}
    \toprule
    \multirow{2}{*}{\textbf{ID}} & \multirow{2}{*}{\textbf{Model}} & \multirow{2}{*}{\textbf{Decoding type}}  & \multicolumn{2}{c}{\textbf{CER (\%)} $\downarrow$}   & \multirow{2}{*}{\textbf{Speed up} $\uparrow$} \\
    & & & \makecell{Dev} & \makecell{Test}\\
    \midrule
    \textbf{A1} & CTC/AED Zipformer & CTC Greedy Search & 3.98 & 4.19  & - \\
    \textbf{B1} & A1 + 3-gram & Dense & 3.76 &  3.99 & 1.18 $\times$  \\
    \textbf{B2} & A1 + 5-gram & Dense & 3.73 & 3.94 & 1.00 $\times$ \\
    \textbf{B3} & A1 + 7-gram  & Dense & 3.73 & 3.95 & 0.91 $\times$ \\
    \midrule
    \textbf{C1} & B2 + LSD & 0.90 \textit{blank} threshold & 3.87 & 4.06 & 2.82 $\times$ \\
    \textbf{C2} & B2 + LSD  & 0.95 \textit{blank} threshold & 3.74 & 3.95 & 1.73 $\times$ \\
    \textbf{C3} & B2 + LSD & 0.99 \textit{blank} threshold & 3.73 & 3.95 & 1.26 $\times$ \\
    \midrule
    \textbf{D1} & B2 + Speech-LLaMA & Averaging &  3.89 & 4.11 & 2.77 $\times$ \\
    \textbf{E1} & B2 + PolyVoice & Discarding &6.33 & 6.79 & 3.23 $\times$ \\
    \midrule
    \textbf{F1} & B2 + Left SWD & $\mathrm{\{-1, 0\}}$ & 3.85 & 4.07 & \textbf{2.50} $\times$  \\
    \textbf{F2} & B2 + Left SWD & $\mathrm{\{-2, -1, 0\}}$ &3.80 & 3.99 & 1.85 $\times$ \\
    \textbf{F3} & B2 + Left SWD &  $\mathrm{\{-3, -2, -1, 0\}}$ & 3.81&3.99& 1.58 $\times$ \\
    \textbf{F4} & B2 + Right SWD & $\mathrm{\{0, 1\}}$ & 3.81&3.99 & 2.35 $\times$ \\
    \textbf{F5} & B2 + Right SWD & $\mathrm{\{0, 1, 2\}}$ & 3.81 & 4.01& 1.87 $\times$ \\
    \textbf{F6} & B2 + Right SWD & $\mathrm{\{0, 1, 2, 3\}}$ &3.84 & 4.05& 1.67 $\times$ \\
    \textbf{F7} & B2 + Both SWD & $\mathrm{\{0, \pm 1\}}$ &3.71&3.92& 2.06 $\times$ \\
    \textbf{F8} & B2 + Both SWD & $\mathrm{\{0, \pm 1, \pm 2\}}$ & \textbf{3.69} &\textbf{3.89}& 1.76 $\times$ \\
    \textbf{F9} & B2 + Both SWD & $\mathrm{\{0,  \pm 1, \pm 2, \pm 3\}}$ &3.74 & 3.94& 1.48 $\times$ \\
    \bottomrule
    \end{tabular}}
    \vspace{-3mm}
\end{table*}

\subsection{Experimental results}
\subsubsection{Benchmark modeling}
To rigorously validate the efficacy of the proposed SWD algorithm in both recognition accuracy and inference speed, we first establish a baseline model, referred to as A1, using a CTC/AED architecture based on zipformer. As shown in Table 1, we perform experiments on the AISHELL-1 dataset and when decoding with greedy search by CTC only, our implemented model achieves SOTA results of 3.98\% and 4.19\% CER on the validation and test sets, respectively. The subsequent experiments are builded on our SOTA model, further substantiate the effectiveness and persuasiveness of the proposed method.

\begin{table}[!t]
    \vspace{1mm}
    \caption{Experiment results on the In-House dataset. }
    \centering
    \renewcommand{\arraystretch}{1.0}
    \label{table3}
    \setlength{\tabcolsep}{2mm}{
    \begin{tabular}{ccccc}
    \toprule
    \textbf{ID} & \textbf{Model} &  \textbf{\makecell{Decoding \\ type}} & \textbf{\makecell{Test  CER(\%) $\downarrow$}} & \textbf{Speed up} $\uparrow$ \\
    \midrule
    \textbf{G1} & A1-XL & Dense & 2.15 & 1.00 $\times$ \\
    \textbf{G2} & G1 + SWD & $\mathrm{\{0, \pm 1\}}$ & \textbf{2.09} &  \textbf{2.17} $\times$\\
    \textbf{G3} & G1 + SWD & $\mathrm{\{0, \pm 1, \pm 2\}}$  & 2.11 & 1.93 $\times$ \\
    \bottomrule
    \end{tabular}}
    \vspace{-4mm}
\end{table}

\subsubsection{Experiment of the AISHELL-1 dataset.}
Table. \ref{tabel2} presents a comparative analysis of the performance of the model when incorporating our proposed SWD algorithm, benchmarked against other prevalent methods in the AISHELL-1 dataset. In the five sets of experiments, labeled C, D, and E, we explore various strategies inspired by methods from prior work on handling CTC posteriori probabilities. Specifically, experiments C1 through C3 focus on evaluating the impact of a label synchronization-based algorithm when integrated with the TLG decoding graph. In these experiments, the probability thresholds of 0.90, 0.95, and 0.99 in the \textit{Decoding type} field represent conditions where frames with \textit{blank} probabilities exceeding these thresholds are discarded, while the remaining frames are preserved for subsequent decoding procedure. The LSD algorithm operates under the hypothesis that as the frame discard threshold increases, recognition accuracy will approach that of the vanilla dense system, but with impaired inference speed. The results from experiments C1 through C3 demonstrate that increasing the discard threshold to 0.95 yields a significant 1.73x increase in inference speed without a notable loss in accuracy. Moreover, experiments D and E involve averaging the blank frames between the neighbouring non-blank frames and discarding all the blank frames, respectively. Although these methods do enhance inference speed, they come with a more substantial trade-off in recognition accuracy.

Experiments F1 through F9 present the results of applying the SWD algorithm proposed in this work. The notation $\mathrm{\{-1, 0\}}$ in F1 represents a scenario where only the left neighbor of the CTC spike is spliced, while F1 through F3 denote the progressively increasing in the range of neighboring frames on the left side of the splice. Experiments F4 through F6 focus on splicing the right neighbor, and experiments F7 through F9 explore the effects of simultaneously splicing both the left and right neighbor spikes. The experimental findings reveal that there are no substantial differences in recognition accuracy and inference speed between splicing only the left neighbor and splicing only the right neighbor. However, compared to E1, the discarding strategy will result in a huge accuracy degradation, this suggests that the left and right neighbors of non-blank spikes carry similar semantic information which is essential to maintain the system performance. Meanwhile, simultaneous splicing of both left and right neighbors yields a 7.29\% and 7.16\% relative reduction in CER in the validation and test sets compared to A1, respectively, along with a 1.76-times improvement in inference speed compared to the B2 system. These results further corroborate the effectiveness of the SWD algorithm in improving both the accuracy and inference speed of the model, demonstrating its potential for practical applications.

\subsubsection{Experiment of the In-House dataset.}
To further substantiate the effectiveness of the SWD algorithm on large-scale datasets, we train a 5-gram TLG model with 1.5 billion internal textual dialog samples, coupled with a 316.7 million parameter acoustic model. Remarkably, as the volume of training data and the model's parameter count increase, the SWD algorithm demonstrates that only one frame adjacent to both the left and right of each spike is sufficient to surpass the performance of the baseline model, even surpassing the performance of adding more frames to both the left and right sides. Furthermore, this approach contributes to a significant enhancement in inference speed, achieving a 2.17-fold improvement over vanilla dense methods.

Our findings suggest that frames adjacent to non-blank spikes encapsulate rich contextual information. By harnessing this neighboring information, SWD not only captures the most critical spikes but also leverages the additional context, leading to a significant boost in performance. This approach, which we describe as $\mathbf{Breaking \ Through \ the \ Spike}$, surpasses traditional dense frame selection methods and simply utilizing CTC spikes strategies. Building on this, we further hypothesize that if arbitrary inserting a \textit{blank} unit near non-blank spikes might achieve accuracy comparable to dense frame selection. These findings introduce a novel and effective paradigm for leveraging CTC outputs, offering new insights for further optimization.

\section{Conclusion}
In this paper, we thoroughly explore the spiking behavior of CTC outputs and propose the conjecture that frames adjacent to non-blank spikes contain semantic information beneficial to the model. Building on this, we present a novel approach to enhance both inference speed and recognition accuracy of CTC-based end-to-end ASR systems named Spike Window Decoding algorithm. By capitalizing on the semantic richness of frames adjacent to non-blank spikes, our method enables a more efficient integration with WFSTs, drastically reducing the number of frames involved in decoding. Additionally, we introduce a weight pushing optimization between the $\mathrm{det}$ and $\mathrm{min}$ steps, improving TLG search efficiency through early pruning. 
The experimental results on both AISHELL-1 dataset and large-scale In-House evaluations confirm that the SWD algorithm can significantly enhance inference speed while even improving recognition accuracy. In future work, we aim to extend the SWD algorithm by formalizing the semantic implications of blank frames, providing a theoretical foundation for speech recognition systems that utilize CTC as the objective function during decoding.

\section{Acknowledgements}
The research is supported by National Natural Science Foundation of China under Grant No. 62306029, Beijing Natural Science Foundation under Grants L233032, Shenzhen Research Institute of Big Data under Grant No. K00120240007.

\bibliographystyle{IEEEtran}
\bibliography{refs}
\end{document}